\newif\ifpdf
\ifx\pdfoutput\undefined
    \pdffalse          % we are not running PDFLaTeX
\else
    \pdfoutput=1       % we are running PDFLaTeX
    \pdftrue
\fi

\documentclass[aps,pra,twocolumn,groupedaddress,floatfix,showpacs,showkeys]{revtex4}

\usepackage{amsmath}
\usepackage{amsthm}
\usepackage{amssymb}

\newcommand{\bra}[1]{\big<#1\big|}
\newcommand{\Bra}[1]{\left<#1\right|}
\newcommand{\ket}[1]{\big|#1\big>}
\newcommand{\Ket}[1]{\left|#1\right>}
\newcommand{\Braket}[2]{\left<#1|#2\right>}

\newcommand{\set}[1]{\{#1\}}
\newcommand{\Set}[1]{\left\{#1\right\}}

\newcommand{\abs}[1]{|#1|}
\newcommand{\Abs}[1]{\left|#1\right|}

\newcommand{\intervalco}[2]{[#1,\, #2)}

\newtheorem{defi}{Definition}
\newtheorem{theo}[defi]{Theorem}
\newtheorem{lemm}[defi]{Lemma}

\input{Qcircuit}

\usepackage[matrix,frame,arrow]{xypic}

\ifpdf
\usepackage[pdftex]{graphicx}
\else
\usepackage[dvips]{graphicx}
\fi

\newcommand{\myfigwidth}[5]{
  \begin{figure}[!t!b!h!]
    \begin{center}
      \ifpdf
      \includegraphics[width=\columnwidth]{#1}
      \else
      \includegraphics[width=\columnwidth]{#2}
      \fi 
      \caption[#5]{#3}
      \label{#4}
    \end{center}
  \end{figure}
}

\ifpdf
\usepackage[pdftex]{hyperref}
\fi

\begin{document}

\title{A Lower Bound for Quantum Phase Estimation}

\date{\today}

\author{Arvid J. Bessen}
\email[]{bessen@cs.columbia.edu}
\affiliation{Columbia University\\
Department of Computer Science}

\begin{abstract}
  We obtain a query lower bound for quantum algorithms solving the phase estimation problem.
  Our analysis generalizes existing lower bound approaches to the case where the oracle $Q$ is given by controlled powers $Q^p$ of $Q$, as it is for example in Shor's order finding algorithm. % \cite{sho-94}.
  In this setting we will prove a $\Omega( \log 1/\epsilon )$ lower bound for the number of applications of $Q^{p_1}, Q^{p_2}, \ldots$.
  This bound is tight due to a matching upper bound.
  We obtain the lower bound using a new technique based on frequency analysis.
\end{abstract}



\maketitle

\section{Introduction}

We study lower bounds for the phase estimation problem.
In this problem we are given a unitary transformation $Q$ as a black-box and we know that $\Ket{q}$ is an eigenvector of $Q$, i.e.
\begin{equation}\label{eqn:Q-phase}
  Q \Ket{q} = e^{2 \pi i \varphi} \Ket{q}, \ \ \varphi \in \intervalco{0}{1}.
\end{equation}
We want to determine the phase $\varphi$ up to precision $\epsilon$.

The quantum phase estimation algorithm approximates $\varphi$ given $\Ket{q}$ and is the main building block in Shor's factoring algorithm, the counting algorithm, and the eigenvalue estimation algorithm \cite{sho-94,nie-chu-00,bra-hoy-tap-98,bra-hoy-mos-tap-00,abr-llo-98,jak-pap-03,pap-woz-04}.

The main element of this algorithm are controlled powers of $Q$, which we define as follows.
Let $Q$ be a $t$ qubit unitary transformation and $\Ket{\psi}$ an arbitrary $t$ qubit state.
For $l=1,\ldots,c$, and $p \in \mathbb{N}$ we define the $c+t$ qubit transformation
\begin{equation}\label{eqn:Wtilde}
  W_{l}^{p}(Q)
  \Ket{x_1 ... x_c} \Ket{\psi}
  =
  \begin{cases}
    \Ket{x_1 ... x_c} \Ket{\psi} & x_l = 0 \\
    \Ket{x_1 ... x_c} Q^p \Ket{\psi} & x_l = 1
  \end{cases}.
\end{equation}
We call $W_{l}^{p}(Q)$ a (controlled) \emph{power query}.
If the transformation $Q$ is clear from the context, we will just write $W_l^p = W_l^p(Q)$.

\begin{figure}[!tbhp!]
  \begin{equation*}
    \Qcircuit @C=4.5pt @R=5pt {
      \Ket{0} \arr & \gate{H^{\otimes T}} & \multigate{1}{W_T^{2^0}} & \multigate{1}{W_{T-1}^{2^1}} & \ldots \arr & \multigate{1}{W_1^{2^{T-1}}} & \gate{\mathcal{F}_{2^T}^{-1}} & \Ket{\widetilde{\varphi}}\\
      \Ket{q} \arr & \wire                & \even{W_T^{2^0}}         & \even{W_{T-1}^{2^1}}         & \ldots \arr & \even{W_1^{2^{T-1}}}         & \wire                         & \Ket{q}\\
    }
  \end{equation*}
  \caption{The quantum phase estimation algorithm in power query notation\label{fig:phase-est-algo-pq}.
    $H$ is the Hadamard gate, $W_l^p$ a power query as in equation (\ref{eqn:Wtilde}), and
    $\mathcal{F}_{2^T}^{-1}$ is the inverse quantum Fourier transform on $T$ qubits.
}
\end{figure}
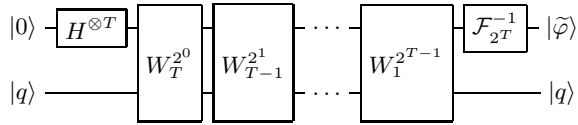

This notation allows us to write the phase estimation algorithm
in a compact form, see figure \ref{fig:phase-est-algo-pq}.
The algorithm returns an approximation $\widetilde{\varphi}$
of the phase $\varphi$ of $\Ket{q}$.

It is well known that $T = \mathcal{O} (\log \epsilon^{-1})$ power queries suffice to approximate $\varphi$ up to $\epsilon$.
In this paper we study whether it is possible to improve on the performance of the phase estimation procedure, i.e., we ask what is the minimal number of applications of $W_l^p$ to estimate $\varphi$ up to $\epsilon$.
\begin{theo}\label{theo:lower-bound-cap}
  Any quantum algorithm estimating the phase $\varphi$ of an eigenvector $\Ket{q}$ of matrices $Q$ up to precision $\epsilon$, with $Q$ from the class
  \begin{multline}\label{eqn:mathcalQ}
    \mathcal{Q}_{\Ket{q},t} = \big\{ Q : Q \text{ is a unitary } t\text{ qubit transform}, \\
    \Ket{q} \text{ is an eigenvector of } Q \big\}.
  \end{multline}
  has to use $\Omega ( \log \frac{1}{\epsilon} )$ power queries.
\end{theo}
We prove Theorem \ref{theo:lower-bound-cap} in section \ref{sec:proofs}.

The query cost of the phase estimation algorithm may be given by counting each application of $W_l^p(Q)$ as $p$ applications of $Q$, i.e. the query cost of this algorithm is
\begin{equation*}
  1 + 2 + 4 + \ldots + 2^{T-1} = 2^T - 1.
\end{equation*}
For certain problems, like order-finding, it is possible to exploit some knowledge about $Q$.
Here $Q \Ket{y} = \Ket{x y \mod N}$ for a certain fixed $x$, and therefore
\begin{equation*}
  Q^{2^j} \ket{y} = \ket{ x^{2^j} y \mod N}
  =  \ket{ \left(x^{2^{j-1}}\right)^2 y \mod N}
\end{equation*}
is easy to compute by repeated squaring and modular multiplication, see e.g. \cite{nie-chu-00}.
In this case we can use power queries $W_l^{2^k}$ with essentially the same cost as an application of $Q$.
Thus we can execute the phase estimation algorithm with query cost $T$ and have exponential speedup for the query cost: from $2^T-1$ to $T$.
Let us stress that this speedup only applies if the cost for computing $Q^{2^j}$ is similar to that for computing $Q$.

\section{Prior work}

Quantum query complexity has been important in quantum computing since Gro\-ver's search algorithm, which is provably superior to classical algorithms \cite{gro-96a,boy-bra-hoy-tap-98} in the number of queries.
The first lower bound result was given in \cite{ben-ber-bra-vaz-97}, which used an adversary argument.

Our lower bound approach is based on the ideas of the ``polynomial approach'' of Beals et. al., \cite{bea-buh-cle-mos-wol-98, nay-wu-99,aar-01,shi-02}.
Other approaches include the quantum adversary argument and its generalizations \cite{amb-00,lap-mag-03,bar-sak-sze-03}.

These approaches only cover problems concerning Boo\-le\-an functions, so we have to extend them to numerical problems.
This has been done through extensions of the polynomial method \cite{hei-01,bes-04} and has been applied widely to integration \cite{nov-01,hei-01b,hei-kwa-woz-03}, path integration \cite{tra-woz-01}, approximation \cite{nov-slo-woz-02,hei-03a,hei-03b}, and eigenvalue estimation \cite{jak-pap-03,pap-woz-04}.

In this paper we apply the approach of \cite{bes-04} to the phase estimation problem.
Instead of using a maximum degree argument, which is not applicable in the case of arbitrary powers, we will develop a new lower bound technique based on frequency analysis.

\section{Quantum Algorithms with Power Queries for Phase Estimation}

We would like to derive lower bounds for any quantum algorithm with power queries that estimates the phase of a matrix $Q$ for a given eigenvector $\Ket{q}$.
In other words the set of allowed inputs for our problem is
\begin{multline*}
  \mathcal{Q}_{\Ket{q},t} = \big\{ Q : Q \text{ is a unitary } t\text{ qubit transform}, \\
  \Ket{q} \text{ is an eigenvector of } Q \big\}.
\end{multline*}

We now give a framework that is general enough to allow us to analyze any algorithm with power queries $W_l^p(Q)$ that solves this problem.
The most general algorithm will be of the following form:
\begin{equation}\label{eqn:cap-query-algo}
  \ket{\psi^{(T)}(Q)}
  = U_{T} W_{l_{T}}^{p_{T}}(Q) U_{T - 1} \ldots W_{l_1}^{ p_1}(Q) U_0 \ket{\psi^{(0)}}.
\end{equation}
Here the $U_0$, $U_1$, $\ldots$, $U_{T}$ are arbitrary but fixed $c+t$ qubit unitary transformations and $\ket{\psi^{(0)}}$ a fixed $c+t$ qubit state, for example $\Ket{0}\Ket{q}$.
In our analysis we neglect the cost to implement the $U_j$ or to prepare $\ket{\psi^{(0)}}$.

The varying parts of algorithm (\ref{eqn:cap-query-algo}) are the $W_{l_j}^{p_j} = W_{l_j}^{p_j}(Q)$: power queries of $Q$ for $p_j \in \mathbb{N}$, $l_j = 1, \ldots, c$, and $c \in \mathbb{N}$ arbitrary.
A measurement of the final state
$\ket{\psi^{(T)}(Q)}$
in the standard basis
yields a state $\Ket{k}$, $k=0,\ldots,2^{c+t}-1$, with probability $p_{k, Q}$, from which we get a solution $\widetilde{\varphi}(k) \in \intervalco{0}{1}$.
If for all $Q \in \mathcal{Q}_{\Ket{q},t}$
the probability to get an $\epsilon$-estimate to the correct phase $\varphi$ of $Q$
\begin{equation}\label{eqn:prob-condition}
  \sum_{ k : \| \varphi - \widetilde{\varphi}(k) \| < \epsilon} p_{k, Q} \geq \frac{3}{4},
\end{equation}
then the algorithm $(\ref{eqn:cap-query-algo})$ solves the phase estimation problem to within $\epsilon$ with probability $\frac{3}{4}$ in $T$ power queries.

We are interested in the smallest number $T$ such that a quantum power query algorithm of form (\ref{eqn:cap-query-algo}) fulfills condition (\ref{eqn:prob-condition}) for all $Q \in \mathcal{Q}_{\Ket{q},t}$.

\section{General controlled arbitrary power queries}\label{sec:proofs}

We consider arbitrary powers $p_1$, $\ldots$, $p_{T}$.
This requires us to introduce a new proof technique.
To illustrate the idea consider the phase estimation algorithm, performed as in figure \ref{fig:phase-est-algo-pq} with $c=T=3$ control qubits.
\begin{equation*}
  (\mathcal{F}_{2^3}^{-1} \otimes I) W_{1}^{4} W_{2}^{2} W_{3}^{1} (H^{\otimes 3} \otimes I) \ket{0} \Ket{q}.
\end{equation*}
Let us trace through each of the steps in this algorithm (we neglect normalization factors).
\begin{enumerate}
\item $(\Ket{0} + \Ket{1} + \Ket{2} + \Ket{3} + \ldots + \Ket{7})\Ket{q}$
\item $(\Ket{0} + e^{2 \pi i \varphi} \Ket{1} + \Ket{2} + e^{2 \pi i \varphi} \Ket{3} + \ldots + e^{2 \pi i \varphi} \Ket{7})\Ket{q}$
\item $(\Ket{0} + e^{2 \pi i \varphi} \Ket{1} + e^{2 \pi i 2 \varphi} \Ket{2} + \ldots + e^{2 \pi i 3 \varphi} \Ket{7})\Ket{q}$

  The possible multiplicities $j$ of $\varphi$ in the coefficients $e^{2 \pi i j \varphi}$ are
  \begin{equation*}
    \mathcal{J}_2 = \set{ 0, p_1, p_2, p_1+p_2 } = \set{0, 1, 2, 3}.
  \end{equation*}
\item $(\Ket{0} + e^{2 \pi i \varphi} \Ket{1} + e^{2 \pi i 2 \varphi} \Ket{2} + \ldots + e^{2 \pi i 7 \varphi} \Ket{7})\Ket{q}$

  The possible multiplicities after this step are
  \begin{equation*}
    \mathcal{J}_{3} = \set{ j, j+p_{3} : j \in \mathcal{J}_2} = \set{0, 1, \ldots, 7}.
  \end{equation*}
\end{enumerate}

The final step, the inverse Fourier transform, does not depend on $\varphi$.
It also does not change the possible multiplicities of $\varphi$, but just creates linear combinations of them.
Consider, e.g., the coefficient of the state $\Ket{2}$:
\begin{equation*}
  \sum_{j=0}^{7} e^{- 2 \pi i 2 j/8} e^{2 \pi i j \varphi} \Ket{2} \Ket{q}
  = \sum_{j=0}^{7} e^{2 \pi i j (\varphi-1/4)} \Ket{2} \Ket{q},
\end{equation*}
which gives the probability $p_2 (\varphi)$ of measuring $\Ket{2}$:
\begin{equation*}
  p_2 (\varphi)
  =
  \Big|
  \sum_{j=0}^{7} e^{2 \pi i j (\varphi-1/4)}
  \Big|^2
  =
  \sum_{j,l=0}^{7} e^{2 \pi i (j-l) (\varphi-1/4)},
\end{equation*}
which is plotted in figure \ref{fig:probOnPhaseKet2}.%
\myfigwidth{probOnPhaseKet2.pdf}{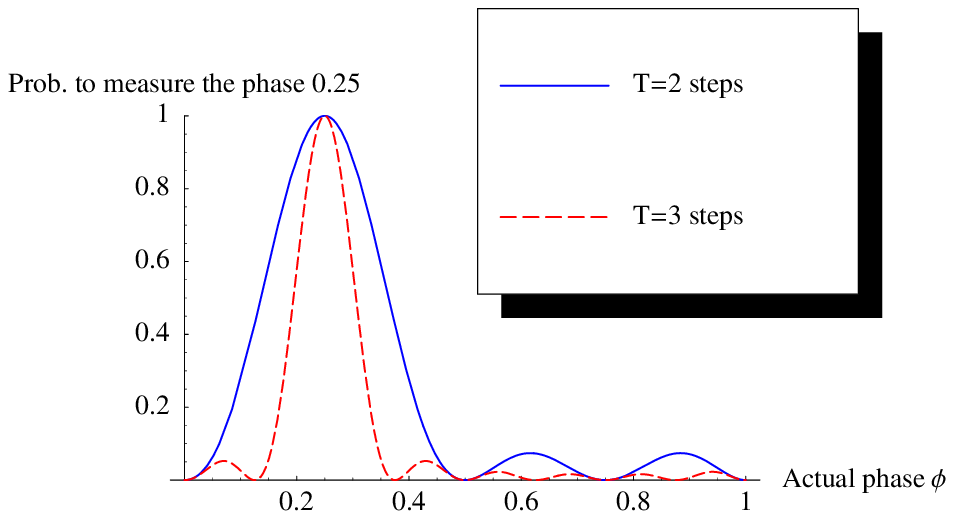}{(Color online) The probability of measuring the state $\Ket{2}$ depending on $\varphi$ for the algorithm depicted in figure \ref{fig:phase-est-algo-pq} with $T=2$ and $T=3$.}{fig:probOnPhaseKet2}{}
Figure \ref{fig:probOnPhaseKet2} shows that the probability that $\Ket{2}$ is measured is high if $\varphi$ is close to $0.25$, which is the value represented by $\Ket{2}$.

The figure indicates that the width of this probability peak depends on the frequencies present in the probability function: higher frequencies allow sharper peaks.
The goal of this paper is to prove that every halving of the width of the probability peak requires one additional step of the algorithm.

The proof consists of three steps.
The first is to quantify the influence of each additional application of $W_l^p$ on the frequencies present in the probability function (Theorem \ref{theo:trig-poly-ap}).
Now consider a probability function as in figure \ref{fig:probOnPhaseKet2}.
It must have a high peak $\geq 3/4$ with small width $\epsilon$ and should be close to zero everywhere else.
We will show that such a function requires a large range of frequencies to be present (lemma \ref{lemm:all-frequencies}).
Finally we will show an upper bound on the number of frequencies that a power query algorithm can generate with $T$ power queries, which proves Theorem \ref{theo:lower-bound-cap}.

In Theorem \ref{theo:trig-poly-ap} consider a general $Q$, which acts on $t$ qubits and therefore has $2^t$ eigenvectors $\Ket{\psi_s}$ with eigenvalues $e^{2 \pi i \varphi_s}$.
We assume that the eigenvectors $\Ket{\psi_s}$ are fixed, but that the eigenvalues change.
We will prove that after $T$ steps only coefficients like 
\begin{equation*}
  \alpha e^{2 \pi i (j_1 \varphi_1 + \ldots + j_{2^t} \varphi_{2^t})}
\end{equation*}
will occur.
Here $(j_1,\ldots,j_{2^t})$ is from the set $\mathcal{J}_T$ defined by the recursion
\begin{multline}\label{eqn:J-recursive}
  \mathcal{J}_{T+1} := \big\{ (j_1, ..., j_{2^t}), (j_1 + p_{T+1}, ..., j_{2^t}), ..., \\
  (j_1, ..., j_{2^t}+p_{T+1}) \, : \, (j_1, ..., j_{2^t}) \in \mathcal{J}_T \big\}
\end{multline}
and $\mathcal{J}_0 = \Set{(0,\ldots,0)}$.% will occur.

\begin{theo}\label{theo:trig-poly-ap}
  Let $Q$ be a unitary operation with eigenvectors $\Ket{\psi_s}$ and corresponding eigenvalues $e^{2 \pi i \varphi_s}$, $s=1, \ldots, 2^t$.
  Let the $\Ket{\psi_s}$ be fixed and vary the phases $\varphi_s \in \intervalco{0}{1}$.
  Any quantum algorithm with power queries $W_l^p = W_l^p(Q)$, fixed unitary transformations $U_j$ and starting state $\ket{\psi^{(0)}}$, can be written as
  \begin{equation}\label{eqn:trig-coeffs}
    U_{T} W_{l_{T}}^{ p_{T}} ... U_1 W_{l_1}^{ p_1} U_0 \ket{\psi^{(0)}}
    = \sum_{k} S_k^{(T)} (\varphi_1, ..., \varphi_{2^t}) \Ket{k}
  \end{equation}
  for all $\varphi_s \in \intervalco{0}{1}$, where the $S_k^{(T)} (\varphi_1, \ldots, \varphi_{2^t})$ are trigonometric polynomials of the following form:
  \begin{multline}\label{eqn:tsnq}
    S_k^{(T)} (\varphi_1, \ldots, \varphi_{2^t}) \\
    = \sum_{(j_{1},\ldots,j_{2^t}) \in \mathcal{J}_{T}}
    \alpha^{(T)}_{k,(j_{1},\ldots,j_{2^t})}
    e^{ 2 \pi i ( j_{1} \varphi_1 + \ldots + j_{2^t} \varphi_{2^t} ) },
  \end{multline}
  with $\alpha^{(T)}_{k,(j_{1},\ldots,j_{2^t})} \in \mathbb{C}$ and $\mathcal{J}_T$ defined by (\ref{eqn:J-recursive}).
\end{theo}
To shorten our proofs we will use the following short-hand notations.
We say that a trigonometric polynomial $S_k^{(T)} (\varphi_1, \ldots, \varphi_{2^t})$ is of powers $\mathcal{J}_T$, indicated by the ${}^{(T)}$ superscript, if it can be written as a sum over $\mathcal{J}_T$ as in (\ref{eqn:tsnq}).
Furthermore we abbreviate the vectors $\vec{j} = (j_1, \ldots, j_{2^t})$, $\vec{\varphi} = (\varphi_1, \ldots, \varphi_{2^t})$, and define the following notation:
\begin{equation*}
  \vec{j} \cdot \vec{\varphi} = j_1 \varphi_1 + \ldots + j_{2^t} \varphi_{2^t}.
\end{equation*}

\begin{proof}[Proof of Theorem \ref{theo:trig-poly-ap}: ]
  To simplify the proof we choose a different basis instead of the standard basis.
  We divide the quantum state into a control part $\Ket{m}$ and an eigenvector part $\Ket{\psi_s}$ and show that we can write
  \begin{equation}\label{eqn:eig-polys}
    U_{T} W_{l_{T}}^{ p_{T}} ... W_{l_1}^{ p_1} U_0 \ket{\psi^{(0)}}
    =
    \sum_{m=0}^{2^c-1} \sum_{s=1}^{2^t}
    \widehat{S}_{m,s}^{(T)}(\vec{\varphi}) \Ket{m, \psi_s},
  \end{equation}
  for trigonometric polynomials of powers $\mathcal{J}_T$,
  \begin{equation}\label{eqn:trig-powers}
    \widehat{S}_{m,s}^{(T)}(\vec{\varphi})
    =
    \sum_{\vec{j} \in \mathcal{J}_T} \widehat{\alpha}_{m,s,\vec{j}}^{(T)} e^{2 \pi i \vec{j} \cdot \vec{\varphi}} .
  \end{equation}
  The proof is by induction on the number of queries $T$.
  For $T=0$ we have no dependence on $\vec{\varphi}$ and $\widehat{S}_{m,s}^{(0)} (\vec{\varphi})$ is just a constant since $Q$ was not applied:
  \begin{equation*}
    \begin{split}
      U_0 \Ket{\psi^{(0)}}
      = &
      \sum_{m=0}^{2^c-1} \sum_{s=1}^{2^t}
      \bra{m,\psi_s} U_0 \ket{\psi^{(0)}}
      \Ket{m, \psi_s} \\
      =: &
      \sum_{m=0}^{2^c-1} \sum_{s=1}^{2^t}
      \widehat{\alpha}_{m,s,(0,\ldots,0)}^{(0)}
      \Ket{m, \psi_s} \\
      = &
      \sum_{m=0}^{2^c-1} \sum_{s=1}^{2^t}
      \sum_{\vec{j} \in \mathcal{J}_0} \widehat{\alpha}_{m,s,\vec{j}}^{(0)} e^{2 \pi i \vec{j} \cdot \vec{\varphi}}
      \Ket{m, \psi_s} .
    \end{split}
  \end{equation*}
%%   with trigonometric polynomials
%%   \begin{equation*}
%%       \widehat{S}_{m,s}^{(0)}(\vec{\varphi})
%%       = % &
%% %      \sum_{\vec{j} \in \Set{(0, \ldots, 0)}} \alpha_{m,s,\vec{j}}^{(0)} e^{2 \pi i \vec{j} \cdot \vec{\varphi}}
%% %      =
%%       \alpha_{m,s,(0,\ldots,0)}^{(0)} % \\
%%       := % &
%%       \bra{m,\psi_s} U_0 \ket{\psi^{(0)}}
%%     \end{split}
%%   \end{equation*}
%%   of powers
%%   $\mathcal{J}_0 = \Set{(0, \ldots, 0)}$.

  Let now $T$ be arbitrary and let equations (\ref{eqn:eig-polys}) and (\ref{eqn:trig-powers}) hold.
  If we apply $W_{l_{T+1}}^{ p_{T+1}}$ to (\ref{eqn:eig-polys}), only those states $\Ket{m, \psi_s}$ are affected for which the $l_{T+1}$-th control bit is set, i.e. $m_{l_{T+1}} = 1$.
  For these states we get
  \begin{equation*}
    W_{l_{T+1}}^{ p_{T+1}}
    \Ket{m, \psi_s}
    =
    \Ket{m} Q^{p_{T+1}} \Ket{\psi_s} \\
    =
    \Ket{m}
    e^{ 2 \pi i p_{T+1} \varphi_s}
    \Ket{\psi_s}
  \end{equation*}
  and therefore the coefficient of $\Ket{m, \psi_s}$ changes to
  \begin{multline*}
    \widehat{S}_{m,s}^{(T)} (\vec{\varphi})
    e^{ 2 \pi i p_{T+1} \varphi_s}
    =
    \sum_{\vec{j}\in \mathcal{J}_{T}}
    \widehat{\alpha}^{(T)}_{m,s,\vec{j}}
    e^{ 2 \pi i \vec{j} \cdot \vec{\varphi} }
    e^{ 2 \pi i p_{T+1} \varphi_s} \\
    =
    \sum_{\vec{j}\in \mathcal{J}_{T}}
    \widehat{\alpha}^{(T)}_{m,s,\vec{j}}
    e^{ 2 \pi i (j_1 \varphi_1 + \ldots + (j_s+p_{T+1}) \varphi_s + \ldots j_{2^t} \varphi_{2^t})}.
  \end{multline*}
  But this is a trigonometric polynomial of powers $\mathcal{J}_{T+1}$ (recall eqn. (\ref{eqn:J-recursive})), and we can write it as
  \begin{equation*}
    \widetilde{S}_{m,s}^{(T+1)} (\vec{\varphi})
    =
    \sum_{\vec{j} \in \mathcal{J}_{T+1}}
    \widetilde{\alpha}^{(T+1)}_{m,s,\vec{j}}
    e^{ 2 \pi i \vec{j} \cdot \vec{\varphi} }
%%     =
%%     \widehat{S}_{m,s}^{(T)} (\vec{\varphi})
%%     e^{ 2 \pi i p_{T+1} \varphi_s}
  \end{equation*}
  if we define the coefficients $\widetilde{\alpha}^{(T+1)}_{m,s,\vec{j}}$ properly:
  \begin{equation*}
    \widetilde{\alpha}^{(T+1)}_{m,s,(j_{1},\ldots, j_s, \ldots ,j_{2^t})}
    =
    \widehat{\alpha}^{(T)}_{m,s,(j_{1},\ldots, j_s - p_{T+1}, \ldots ,j_{2^t})}
  \end{equation*}
  for $p_{T+1} \leq j_s$ and $0$ otherwise.

  Finally we define $\widetilde{S}_{m, s}^{(T+1)} (\vec{\varphi}) := \widehat{S}_{m, s}^{(T)} (\vec{\varphi})$ for the states for which the control bit is not set ($m_{l_{t+1}} = 0$) and we can write
  \begin{equation*}
    W_{l_{T+1}}^{ p_{T+1}} U_T \ldots W_{l_1}^{p_1} U_0 \ket{\psi^{(0)}}
    =
    \sum_{m=0}^{2^{c}-1}
    \sum_{s=1}^{2^t}
    \widetilde{S}_{m, s}^{(T+1)} (\vec{\varphi})
    \Ket{m, \psi_s}.
  \end{equation*}

  Now we use that the transformation $U_{T+1}$ and the eigenvectors $\Ket{\psi_s}$ are fixed for all algorithms we consider and define
  $
  u_{m,s;n,t}^{(T+1)}
  =
  \Bra{m,\psi_s}
  U_{T+1}
  \Ket{n, \psi_{t}}
  $.
  Then
  \begin{multline}\label{eqn:unitary-step1}
    U_{T+1}
    \sum_{n, t}
    \widetilde{S}_{n,t}^{(T+1)}(\vec{\varphi})
    \Ket{n, \psi_{t}} \\
    =
    \sum_{m, s, n, t}
    \widetilde{S}_{n,t}^{(T+1)}(\vec{\varphi})
    u_{m,s;n,t}^{(T+1)}
    \Ket{m,\psi_s} ,
  \end{multline}
  which gives the following coefficient for $\Ket{m, \psi_s}$:
  \begin{equation}\label{eqn:unitary-step2}
    \begin{split}
%%       \sum_{n, t}
%%       \widetilde{S}_{n,t}^{(T+1)}(\vec{\varphi})
%%       u_{m,s;n,t}
%%       =
      &
      \sum_{n, t}
      \sum_{\vec{j} \in \mathcal{J}_{T+1}}
      \widetilde{\alpha}^{(T+1)}_{n,t,\vec{j}}
      e^{ 2 \pi i \vec{j} \cdot \vec{\varphi} }
      u_{m,s;n,t}^{(T+1)} \\
      = &
      \sum_{\vec{j} \in \mathcal{J}_{T+1}}
      \bigg[
	\sum_{n, t}
	u_{m,s;n,t}^{(T+1)}
	\widetilde{\alpha}^{(T+1)}_{n,t,\vec{j}}
	\bigg]
      e^{ 2 \pi i \vec{j} \cdot \vec{\varphi} } \\
      =: &
      \sum_{\vec{j} \in \mathcal{J}_{T+1}}
      \widehat{\alpha}^{(T+1)}_{m,s,\vec{j}}
      e^{ 2 \pi i \vec{j} \cdot \vec{\varphi} }
      =:
      \widehat{S}_{m,s}^{(T+1)} ( \vec{\varphi} ) .
    \end{split}
  \end{equation}
  This completes the induction and establishes equations (\ref{eqn:eig-polys}) and (\ref{eqn:trig-powers}).

  Using the same argumentation as in equations (\ref{eqn:unitary-step1}) and (\ref{eqn:unitary-step2}) we can finally rewrite the state in equation (\ref{eqn:eig-polys}) in the standard basis $\Ket{k} \in \Set{\Ket{0}, \Ket{1}, \ldots, \Ket{2^{c+t}-1}}$ through
  \begin{equation*}
    \alpha^{(T+1)}_{k,\vec{j}}
    =
    \sum_{m, s}
    \Braket{k}{m, \psi_{s}}
    \widehat{\alpha}^{(T+1)}_{m,s,\vec{j}}
  \end{equation*}
  and $S_{k}^{(T)} ( \vec{\varphi} )$ is of the same powers as $\widehat{S}_{m,s}^{(T)} (\vec{\varphi})$, which is of powers $\mathcal{J}_T$.
  This proves equation (\ref{eqn:trig-coeffs}) and (\ref{eqn:tsnq}).
\end{proof}

We now focus on the specific problem of phase estimation.
The next lemma provides us with a necessary condition on the powers $p_1, p_2, \ldots$ such that a quantum algorithm with power queries can solve the phase estimation problem with precision $\epsilon$.

\begin{lemm}\label{lemm:all-frequencies}
  Any quantum algorithm estimating the phase $\varphi$ of an eigenvector $\Ket{q}$ of matrices $Q$ from the class
  \begin{multline*}
    \mathcal{Q}_{\Ket{q},t} = \big\{ Q : Q \text{ is a unitary } t\text{ qubit transform}, \\
    \Ket{q} \text{ is an eigenvector of } Q \big\}.
  \end{multline*}
  up to precision $\epsilon$
  has to use power queries $W_{j_1}^{p_1}$, $W_{j_2}^{p_2}$, $\ldots$, $W_{j_T}^{p_T}$ such that the set
  \begin{equation}\label{eqn:MT}
    \mathcal{M}_T = \big\{
    l - l'
    \, | \, 
    l, l' = \sum_{ k \in K } p_k \, \big| \, K \subseteq \set{ 1, \ldots, T } \big\}
  \end{equation}
  has more than $\Abs{\mathcal{M}_T} \geq \frac{1}{2\epsilon}$ elements.
\end{lemm}
Note that this is both a condition on $p_1, \ldots, p_T$ as well as on $T$: by cleverly choosing the powers $p_1, \ldots, p_T$ we can get away with a smaller $T$, e.g. for $p_j=2^{j-1}$ we get
\begin{equation*}
  \mathcal{M}_T = \Set{ -2^T+1, -2^T+2, \ldots, 2^T-1 },
\end{equation*}
while for the choice $p_j = 1$ we only get
\begin{equation*}
  \mathcal{M}_T = \Set{ -T, -T+1, \ldots, T-1, T }.
\end{equation*}
\begin{proof}[Proof of lemma \ref{lemm:all-frequencies}:]
  For simplicity assume that
  \begin{equation}\label{eqn:Nreqmts}
    2 \epsilon = \frac{1}{N} \text{ for some } N \in \mathbb{N}.
  \end{equation}
  We will analyze the behavior of all possible algorithms for the phase estimation problem on a special subset of $\mathcal{Q}_{\Ket{q},t}$.
  Fix some arbitrary vectors $\Ket{\psi_2}, \ldots , \Ket{\psi_{2^t}}$ such that $\Ket{q}, \Ket{\psi_2}, \ldots, \Ket{\psi_{2^t}}$ form an orthonormal basis and consider the following input:
  \begin{equation}\label{eqn:Qm}
    Q_{r}
    :=
    e^{2 \pi i 2 r \epsilon} \Ket{q} \Bra{q}
    + \sum_{s=2}^{2^t}
    e^{2 \pi i \varphi_s}
    \Ket{\psi_s} \Bra{\psi_s} .
  \end{equation}
  The phase $\varphi$ we are interested in is $\varphi = 2 r \epsilon$ for input $Q_r$.

  Since the difference between the phases of the matrices $Q_r$ is $2 \epsilon$ and we require $\epsilon$ correctness, a measurement will yield states $\Ket{k}$ from the \emph{distinct} sets $B_r$:
  \begin{equation}\label{eqn:Br}
    B_r = \Set{ k : \| 2 r \epsilon - \widetilde{k} \| < \epsilon}.
  \end{equation}
  Depending on the number of qubits we use in our quantum algorithm, the sets $B_r$ can contain one or more states.

  By Theorem \ref{theo:trig-poly-ap} we know that we can write the coefficient of a state $\Ket{k} \in B_r$ after $T$ queries as
  \begin{multline*}
    S_{k}^{(T)} (\varphi, \varphi_2, \ldots, \varphi_{2^t}) \\
    = \sum_{(j_1,\ldots,j_{2^t}) \in \mathcal{J}_{T}}
    \alpha^{(T)}_{k,(j_1,\ldots,j_{2^t})}
    e^{ 2 \pi i ( j_1 \varphi + j_2 \varphi_2 + \ldots + j_{2^t} \varphi_{2^t} ) } .
  \end{multline*}
  In this proof we are only interested in the behavior for the $Q_r$.
  Therefore we can drop the dependence on $\varphi_2, \ldots, \varphi_{2^t}$ and let
  \begin{equation*}
    S_{k}^{(T)} (\varphi)
    := S_{k}^{(T)} (\varphi, \varphi_2, \ldots, \varphi_{2^t})
    = \sum_{l  \in \mathcal{L}_T }
    \beta_{k,l}^{(T)}
    e^{2 \pi i l \varphi},
  \end{equation*}
  where $\mathcal{L}_0 = \Set{0}$ and
  \begin{equation*}
    \begin{split}
      \mathcal{L}_T 
      = &
      \big\{ j_1 \, : \, (j_1, \ldots, j_{2^t}) \in \mathcal{J}_T \big\} \\
      = &
      \big\{ j_1, j_1 + p_T \, : \, (j_1, \ldots, j_{2^t}) \in \mathcal{J}_{T-1} \big\} \\
      = &
      \big\{ \sum_{ k \in K } p_k \, \big| \, K \subseteq \set{ 1, \ldots, T } \big\}
    \end{split}
  \end{equation*}
  and the coefficients (note that $l$ is fixed on the right side)
  \begin{equation*}
    \beta_{k,l}^{(T)}
    = \sum_{(l,j_2,...,j_{2^t}) \in \mathcal{J}_{T}}
    \alpha^{(T)}_{k,(l,j_2,...,j_{2^t})}
    e^{ 2 \pi i ( j_2 \varphi_2 + ... + j_{2^t} \varphi_{2^t} ) }
  \end{equation*}

  The probability $p_{B_r} (\varphi)$ of measuring a state from the set $B_r$ defined in (\ref{eqn:Br}) of all $\epsilon$ approximations to $\varphi = 2 r \epsilon$ is now given by:
  \begin{equation}\label{eqn:pBr}
    \begin{split}
      p_{B_r} (\varphi)
      := & \sum_{k \in B_r} \Abs{S_{k}^{(T)} (\varphi)}^2 \\
      = &
      \sum_{k \in B_r} 
      \sum_{l  \in \mathcal{L}_T }
      \sum_{l'  \in \mathcal{L}_T }
      \beta_{k,l}^{(T)}
      \overline{\beta_{k,l'}^{(T)}}
      e^{ 2 \pi i (l - l') \varphi} \\
      =: &
      \sum_{l \in \mathcal{L}_T }
      \sum_{l' \in \mathcal{L}_T }
      \gamma_{r,l,l'}^{(T)}
      e^{ 2 \pi i (l - l') \varphi} \\
      =: & \sum_{m \in \mathcal{M}_T}
      \eta_{r,m}^{(T)}
      e^{ 2 \pi i m \varphi} \\
    \end{split}
  \end{equation}
  with the set $\mathcal{M}_T$ given by
  \begin{equation*}
    \mathcal{M}_T = \Set{ l - l' \, | \, l, l' \in \mathcal{L}_T }
  \end{equation*}
  and the coefficient
  \begin{equation*}
    \eta_{r,m}^{(T)}
    = \sum_{\substack{l, l' \in \mathcal{L}_T\\ l - l' = m} }
    \gamma_{r,l,l'}^{(T)}
    = \sum_{\substack{l, l' \in \mathcal{L}_T\\ l - l' = m} }
    \sum_{k \in B_r} 
    \overline{\beta}_{k,l}^{(T)}
    \beta_{k,l'}^{(T)} .
  \end{equation*}

  For illustration recall figure \ref{fig:probOnPhaseKet2}, which shows exactly one of these probability functions $p_{B_r}(\varphi)$ and also their highly oscillatory behavior.
  In the case of the phase estimation algorithm, $B_r = \Set{\Ket{r}}$ and the figure shows $p_{B_2}(\varphi)$.

  We apply the Discrete Inverse Fourier Transform to $p_{B_r}(\varphi)$ (evaluated at the points $\varphi = n/N$ for $n=0, \ldots, N-1$) and get for the $k$-th coefficient
  \begin{equation}\label{eqn:prob-dft}
    \begin{split}
      &
      \sum_{n=0}^{N-1} p_{B_r}( n/N ) e^{- 2 \pi i k n / N} \\
      = &
      \sum_{n=0}^{N-1}
      \sum_{m  \in \mathcal{M}_T }
      \eta_{r,m}^{(T)}
      e^{ 2 \pi i (m - k) n / N } \\
      = &
      N
      \sum_{\substack{m  \in \mathcal{M}_T\\m \equiv k \mod N}}
      \eta_{r,m}^{(T)},
    \end{split}
  \end{equation}
  since for $m \not\equiv k \mod N$
  \begin{equation}
    \sum_{n=0}^{N-1}
    e^{ 2 \pi i (m - k) n / N}
    =
    \frac{1 - e^{ 2 \pi i (m - k) N / N}}{1 - e^{ 2 \pi i (m - k) 1 / N}}
    = 0.
  \end{equation}

  We can bound (\ref{eqn:prob-dft})
  by separating the part where a state from $B_r$ is correctly returned from $p_{B_r}(\varphi)$,
  \begin{equation}\label{eqn:prob-lower}
    \begin{split}
      &
      \Big| \sum_{n=0}^{N-1} p_{B_r}( \frac{n}{N} ) e^{- \frac{2 \pi i k n}{N}} \Big| \\
      \geq &
      \Big| p_{B_r}( \frac{r}{N} ) e^{- \frac{2 \pi i k r}{N}} \Big|
      - \Big| \sum_{\substack{n=0\\n \neq r}}^{N-1} p_{B_r}( \frac{n}{N} ) e^{- \frac{2 \pi i k n}{N}} \Big| \\ 
      \geq &
      \frac{3}{4}
      - \sum_{\substack{n=0\\n \neq r}}^{N-1} p_{B_r}( \frac{n}{N} ) ,
    \end{split}
  \end{equation}
  since the probability
  $p_{B_r} (\varphi)$ has to obey
  $p_{B_r} (r /N) \geq \frac{3}{4}$
  (recall the definitions of $p_{B_r}$ and $Q_r$).
  If we knew that for the second term in (\ref{eqn:prob-lower})
  \begin{equation}\label{eqn:prob-balance}
    \sum_{\substack{n=0\\n \neq r}}^{N-1} p_{B_r}( n/N ) < \frac{3}{4}
  \end{equation}
  we could establish that
  \begin{equation*}
    \Abs{\sum_{n=0}^{N-1} p_{B_r}( n/N ) e^{- \frac{2 \pi i k n}{N}}} \\
    =
    \Bigg|
      N
      \sum_{\substack{m  \in \mathcal{M}_T\\m \equiv k \mod N}}
      \eta_{r,m}^{(T)}
    \Bigg|
    > 0.
  \end{equation*}
  We will show that this property, while not necessarily always true, will be true for at least most of the $p_{B_r} (\varphi)$.

  There are $N$ different possible outcome sets $B_0$, $\ldots$, $B_{N-1}$.
  We know that for any $\varphi = n/N$ all mu\-tu\-al\-ly exclusive probabilities of measuring a state from $B_r$ for $r=0, \ldots, N-1$ have to add up to at most $1$:
  \begin{equation*}
    \sum_{r = 0}^{N-1} p_{B_r} (n /N) \leq 1
    \text{ for } n=0, \ldots, N-1,
  \end{equation*}
  Let $R^< \subseteq \Set{0, \ldots, N - 1}$ be the set of all $r$ for which (\ref{eqn:prob-balance})
  holds and $R^\geq$ the set for which it does not.
  $\abs{R^<}$ has to be greater than $1$ since we can split
  \begin{equation*}
    N
    =
    \sum_{n = 0}^{N-1}
    1
    \geq
    \sum_{n = 0}^{N-1}
    \sum_{r = 0}^{N-1}
    p_{B_r} (\frac{n}{N})
  \end{equation*}
  into the following parts:
  \begin{multline*}
    \sum_{r = 0}^{N-1}
    p_{B_r} (\frac{r}{N}) 
    +
    \sum_{r \in R^<}
    \sum_{\substack{n=0\\n \neq r}}^{N-1}
    p_{B_r} (\frac{n}{N})
    +
    \sum_{r \in R^\geq}
    \sum_{\substack{n=0\\n \neq r}}^{N-1}
    p_{B_r} (\frac{n}{N}) \\
    \geq
    \frac{3}{4} N
    + 0 \Abs{R^<} 
    + \frac{3}{4} \Abs{R^\geq} 
    =
    \frac{3}{4} N
    + \frac{3}{4} \Abs{R^\geq}
  \end{multline*}
  and therefore $\Abs{R^\geq} \leq \frac{1}{3} N$ or $\Abs{R^<} \geq \frac{2}{3} N$.

  Thus there is an $r \in R^<$ for which eqn. (\ref{eqn:prob-balance}) holds and
  \begin{equation}\label{eqn:prob-lower-consequence}
    0
    <
    \Bigg|
    N
    \sum_{\substack{m  \in \mathcal{M}_T\\m \equiv k \mod N}}
    \eta_{r,m}^{(T)}
    \Bigg|
    \leq
    N
    \sum_{\substack{m  \in \mathcal{M}_T\\m \equiv k \mod N}}
    \Big|
    \eta_{r,m}^{(T)}
    \Big|
  \end{equation}
  for all $k = 0, \ldots, N-1$.

  This means at least $N$ of the $\eta_{r,m}^{(T)}$ have to be nonzero and thus $p_{B_r} (\varphi)$ from eqn. (\ref{eqn:pBr})
  must have at least $N$ nonzero terms.
  In other words
  \begin{equation}\label{eqn:Mcard-lower-bound}
    \Abs{ \mathcal{M}_T } \geq N = \frac{1}{2 \epsilon},
  \end{equation}
  where we used the definition of $N$ in (\ref{eqn:Nreqmts}).
\end{proof}

Lemma \ref{lemm:all-frequencies} now allows us to give a lower bound for the phase estimation problem.
The numbers of queries $T$ that is needed by any quantum algorithm is $\Omega(\log \frac{1}{\epsilon})$.
\begin{proof}[Proof of Theorem \ref{theo:lower-bound-cap}:]
  From lemma \ref{lemm:all-frequencies} we know that the set $\mathcal{M}_T$ has to have
  $\Abs{\mathcal{M}_T} \geq \frac{1}{2 \epsilon}$ elements.
  We can easily derive an upper bound on $\Abs{ \mathcal{M}_T }$.
  There are at most $2^T$ elements in the set
  \begin{equation*}
    \mathcal{L}_T
    = \big\{ \sum_{ k \in K } p_k \, \big| \, K \subseteq \set{ 1, \ldots, T } \big\}
  \end{equation*}
  and therefore we have
  $%\begin{equation*}
    \Abs{ \mathcal{M}_T }
    \leq \Abs{ \mathcal{L}_T }^2
    \leq (2^{T})^2
    = 2^{2 T}.
  $ %\end{equation*}
  Combining our estimates for $\Abs{\mathcal{M}_T}$ we get
  \begin{equation*}
    2^{2 T} \geq \Abs{\mathcal{M}_T} \geq \frac{1}{2 \epsilon},
  \end{equation*}
  and the number of queries $T$ must grow like %logarithmically in $1/\epsilon$:
  \begin{equation*}
    T \geq \frac{1}{2} \log_2 \frac{1}{2 \epsilon} = \Omega( \log \frac{1}{\epsilon} ) .
  \end{equation*}
\end{proof}

\section{Conclusions and extensions}

In this paper we have obtained lower bounds for quantum algorithms that approximate the phase estimation problem through the new lower bound proof technique of frequency analysis.
These lower bounds match the known upper bounds for phase estimation.

The frequency analysis method can be used to give lower bounds for the Sturm-Liouville eigenvalue problem \cite{pap-woz-04}.
The results also extend to other forms of quantum queries, e.g. the query
\begin{equation*}
  Q_f \Ket{x} \Ket{y} = \Ket{x} \Ket{y \oplus f(x)} = \Ket{x} \Ket{y + f(x) \mod 2^m}.
\end{equation*}
An application of the frequency analysis method to these problems will be the subject of future work.

\begin{acknowledgments}
  The author would like to thank J. Traub, H. Wo{\'z}niakowski, and A. Papageorgiou for inspiring discussions.
  Funding was provided by Columbia University through a Presidential Fellowship.
  This research was supported in part by the National Science Foundation and by the Defense Advanced Research Projects Agency (DARPA) and Air Force Research Laboratory, Air Force Material Command, USAF, under agreement number F30602-01-2-0523.
\end{acknowledgments}

\bibliography{qc}

\end{document}